\documentclass[graybox]{svmult}

\usepackage{mathptmx}       
\usepackage{helvet}         
\usepackage{courier}        
\usepackage{type1cm}        
\usepackage{makeidx}         
\usepackage{graphicx}        
\usepackage{multicol}        
\usepackage[bottom]{footmisc}

\makeindex             

\begin{document}

\title*{Simulating the Synchronizing Behavior of High-Frequency Trading in Multiple Markets}
\author{Benjamin Myers and Austin Gerig}
\institute{Benjamin Myers \at Department of Physics, University of Oxford, Oxford, United Kingdom \email{myers.benjamin.s@gmail.com}
\and Austin Gerig \at CABDyN Complexity Centre, Sa\"{i}d Business School, University of Oxford, Oxford, United Kingdom \email{austin.gerig@sbs.ox.ac.uk}}
%
%
\maketitle

\abstract{Nearly one-half of all trades in financial markets are executed by high-speed autonomous computer programs -- a type of trading often called high-frequency trading (HFT).  Although evidence suggests that HFT increases the efficiency of markets, it is unclear how or why it produces this outcome.  Here we create a simple model to study the impact of HFT on investors who trade similar securities in different markets.  We show that HFT can improve liquidity by allowing more transactions to take place without adversely affecting pricing or volatility.  In the model, HFT synchronizes the prices of the securities, which allows buyers and sellers to find one another across markets and increases the likelihood of competitive orders being filled.}

\section{Introduction}
\label{sec:1}
Financial markets have changed considerably over the last 20 years.  During this time, most exchanges have switched from floor-based to fully electronic trading where orders can be sent to the market and executed with little or no human involvement\cite{MacKenzie}.  As a result, automated trading has flourished.  One particular type of automated trading, known as high-frequency trading (hereafter HFT), has especially grown in size and importance.  HFT exploits short-term price fluctuations and seeks a small profit per transaction many times throughout the day, without taking on significant overnight positions.  Although difficult to determine its true size, most studies estimate that about one-half of all transactions on major exchanges are due to HFT\footnote{Several research firms provide estimates of HFT activity for subscribers; examples are the TABB Group, the Aite Group, and Celent. Publicly, this information is available in articles such as ``The fast and the furious'', Feb. 25, 2012, \emph{The Economist} and ``Superfast traders feel the heat as bourses act'', Mar. 6, 2012, \emph{Financial Times}.}

This study focuses on one particular effect linked to HFT -– the synchronizing of price responses across multiple related securities\cite{Gerig2012, GerigMichayluk2010}.  Figure 1 (taken from a recent article) shows this effect.  Here, to analyze price synchronization in more detail, we simulate two markets where an identical security is traded and compare investor welfare when the prices in these markets are and are not aligned by the actions of HFT.

%
\begin{figure}[htb]
\sidecaption
\includegraphics[scale=.5]{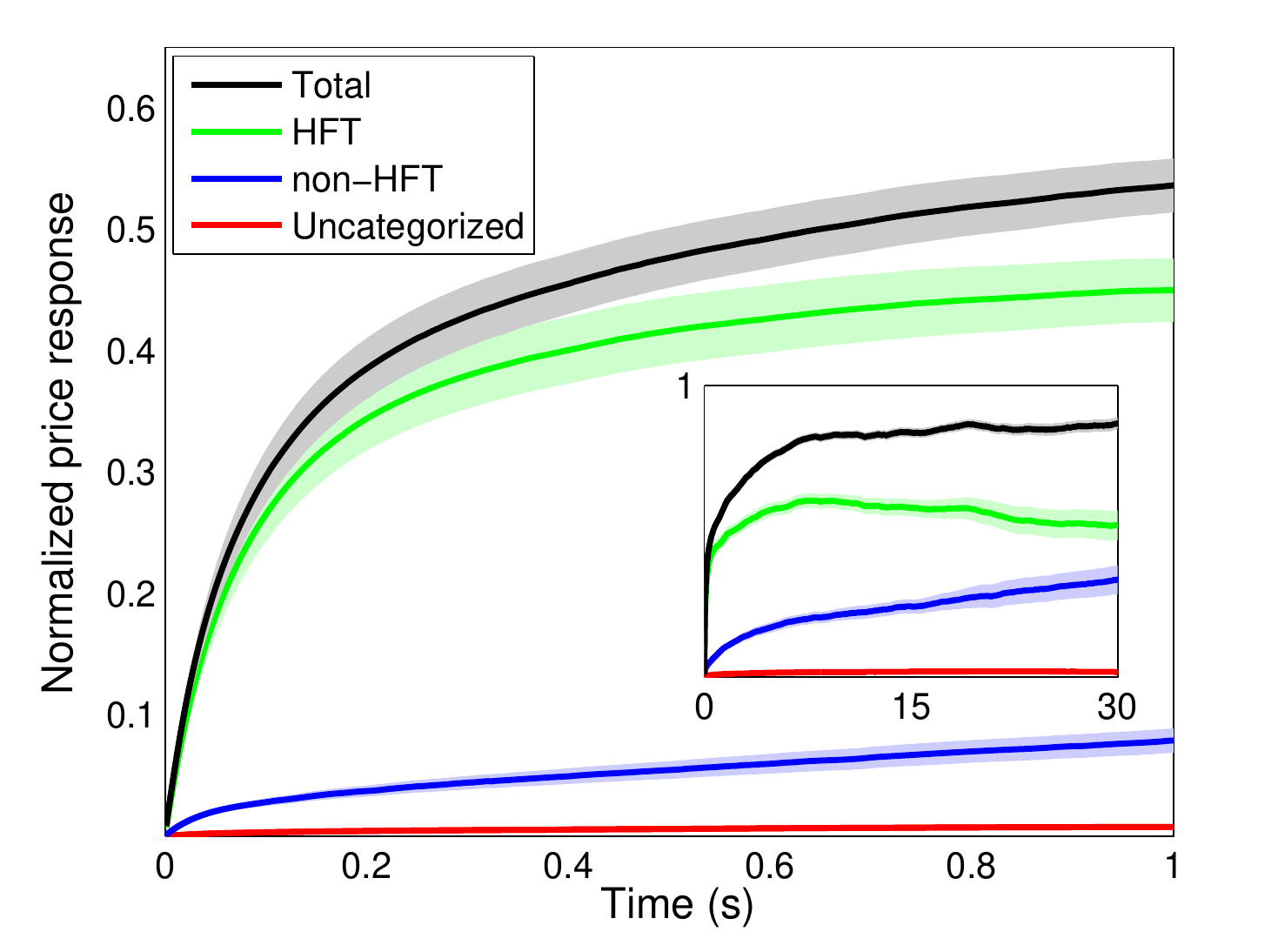}
%
%
\caption{Normalized price response of stock $i$ due to stock $j\neq i$, for 40 US stocks traded on NASDAQ, decomposed into the amount due to HFT activity (green), non-HFT activity (blue) and uncategorized activity (red). Standard errors of the sample means are indicated in the shaded color. Taken from \cite{Gerig2012}.}
\label{fig:1}       
\end{figure}
 
In our simulation, investors are modeled in a zero-intelligence framework\cite{Gode, Farmer}.  This treatment strips out the idiosyncrasies of individuals' behavior and assumes only local interactions are of significance -– investors are only interested in meeting their own specific price expectations and they do not use complex strategies. To consider the effect of HFT, we simulate two zero-intelligence markets where an identical security is traded and allow HFT to connect orders between the two markets when their prices cross. We show that HFT activity (as defined in the model) increases the probability that a typical investor entering the market will transact.  Furthermore HFT activity reduces volatility so that prices are closer to their fundamental value. 

\section{Model}
\label{sec:2}

The model simulates the continuous double auction, a market structure common to most modern exchanges.  Traders submit bids and offers to buy and sell respectively at the best price they are willing to transact at. If prices cross -- a bid meets or exceeds a previous offer, or the converse –- a transaction takes place at the earlier listed price.  If an incoming order is unable to transact with any existing orders, it is placed in the limit order book. This consists of two lists, the bid book and the ask book which contain the previously unfilled orders on the buy and sell side respectively.

At each time step in our model, a random order of unit size is generated.  Orders have equal probability of being a buy or sell and are given a price drawn from a uniform probability distribution with limits 1 and 200.  These hard limits on order prices should not be thought of as boundaries that would exist in real markets, but instead are assumed for simplicity.  In real markets, we would expect participants to place limit orders according to some humped shaped distribution around the equilibrium clearing price (which would change through time).  For simplicity, we assume this humped distribution is a uniform distribution with limits and that the clearing price is constant through time.  Using a dynamic clearing price and/or a different distribution with open limits (such as a Gaussian), although perhaps more realistic, would not change the main results of the paper.

Orders fill the limit order book until a transaction takes place.  When a transaction occurs, all unfilled orders in the limit order book are cleared and the process of generating new orders is started again.  Figure 2(A) illustrates this diagrammatically.

%
\begin{figure}[htb]
\includegraphics[scale=.25]{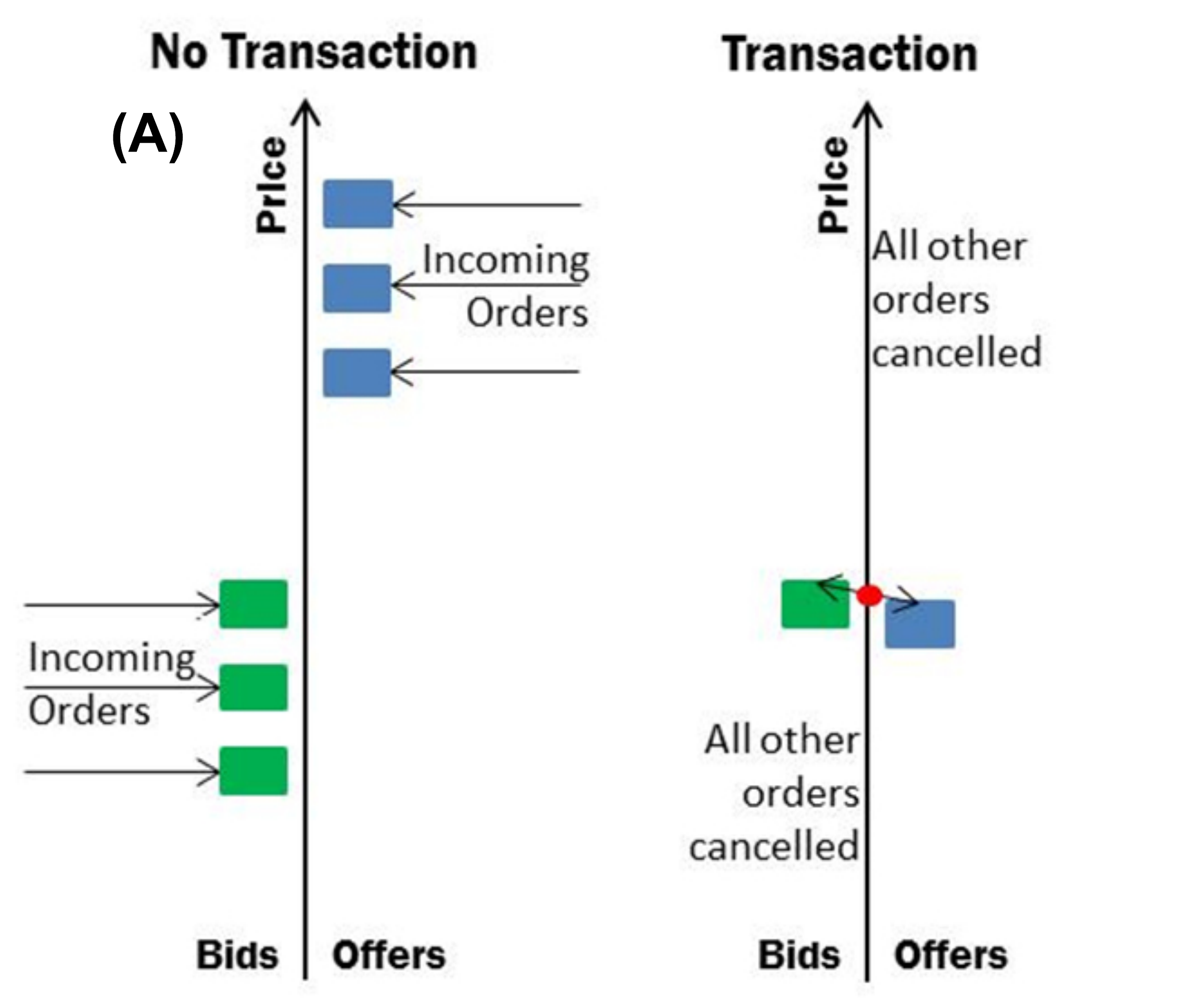}
\includegraphics[scale=.25]{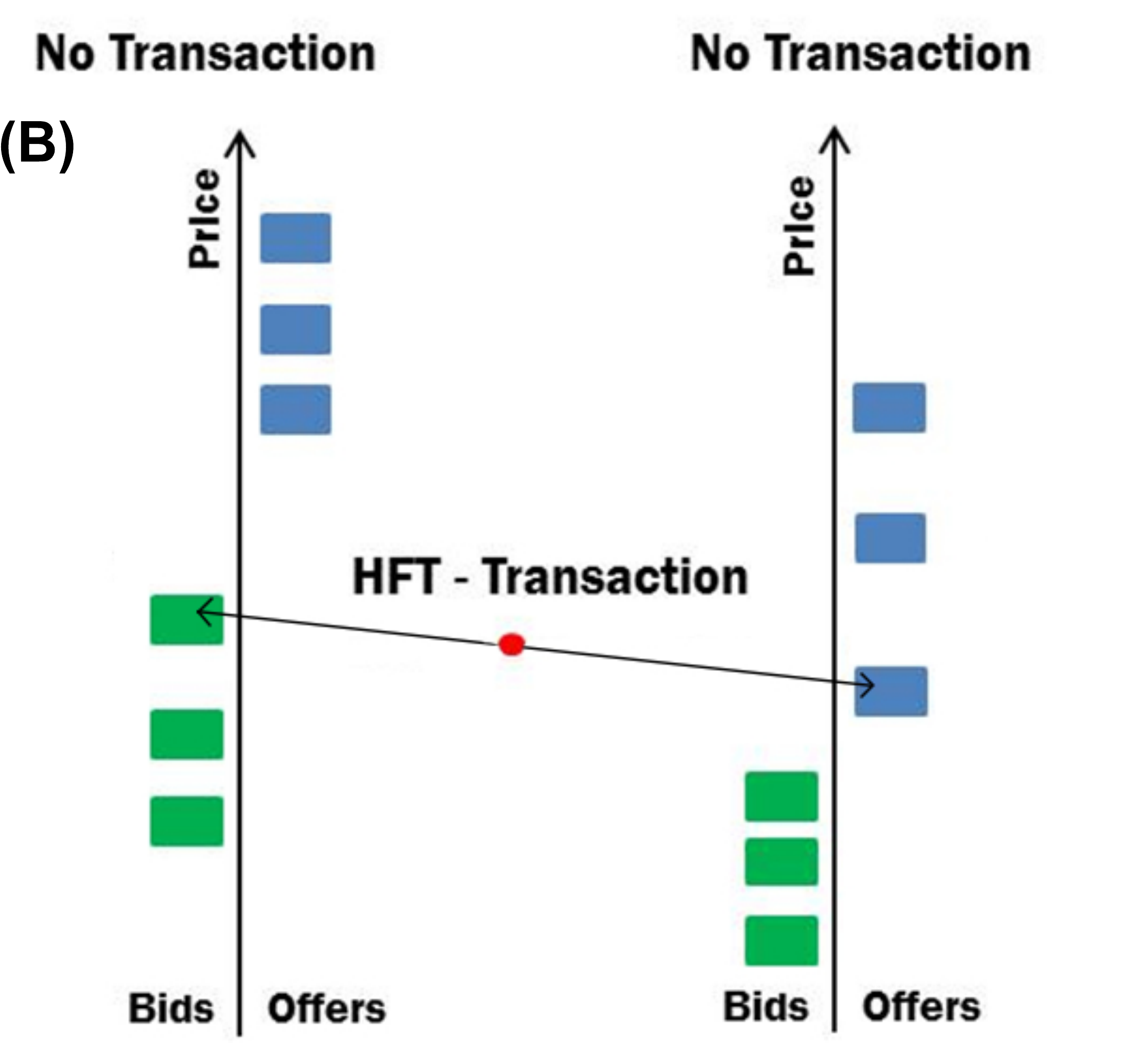}
%
%
\caption{A diagram of the order book in both scenarios modeled. (A) exhibits the market without HFT, at time steps with and without a transaction occurring. Note that transactions can only occur if a bid exceeds an offer. (B) shows the connecting effect HFT. The order book is cleared entirely after each transaction. }
\label{fig:2}       
\end{figure}

The HFT interaction is modeled by running two identical exchanges simultaneously. If transactions are unable to take place on either market in a given time step, but would occur if the two markets were combined, HFT is permitted to transact between the two markets.  Figure 2(B) illustrates this diagrammatically.  We make an idealized assumption that HFT is perfectly competitive so that their profit is zero.  Therefore, the HFT transactions in the two markets take place at the same price, which we set to the midpoint between the bid price and offer price of the two orders in the two markets.

For example, let the bid price on market 1 be denoted by $b_1$ and the ask price on market 2 be denoted by $a_2$, where $b_1 \geq a_2$.  When HFT transacts with these orders, the transactions take place at price $(b_1+a_2)/2$ in both markets.

\section{Results}
\label{sec:3}

When buying or selling a security, investors typically are interested in the following three questions: How likely am I to transact? What price am I likely to receive? How much does this transaction price vary?  We therefore estimate the following three observables in the model both with and without HFT: (1) the probability that a submitted order will result in a transaction, (2) the average transaction price of filled orders, and (3) the volatility of the transaction price of filled orders (the standard deviation of the transaction price).  We run the simulation 100 times for 10000 time steps both with and without HFT, and we record average values of the relevant observables.  The results are shown in Figures 3 and 4 and in Table 1, which we discuss in more detail below.

%
\begin{figure}[htb]
\includegraphics[scale=.33]{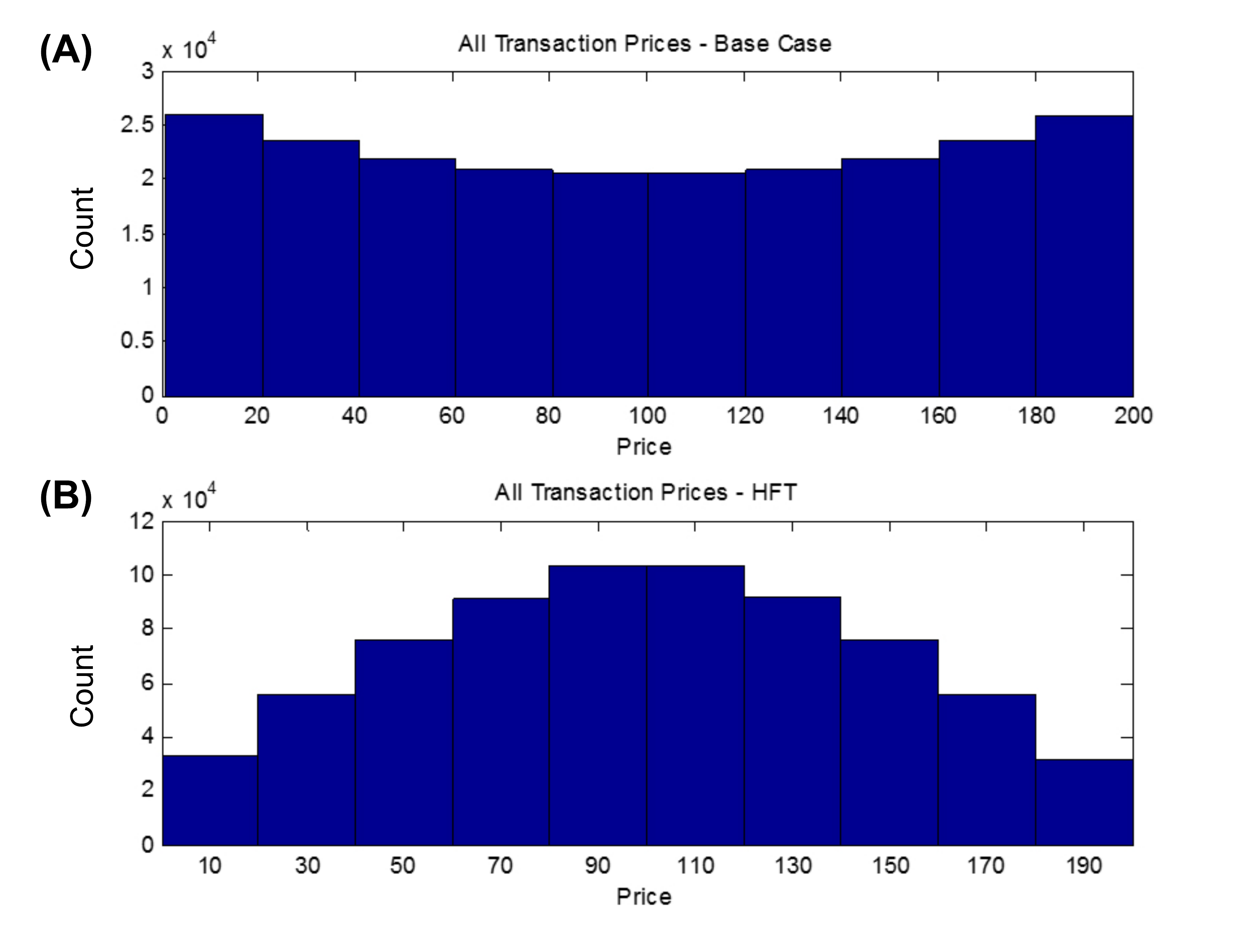}
%
%
\caption{Histogram of transaction prices in the simulation. (A) shows the market without HFT; (B) shows the market with HFT, including transactions over both exchanges. Note that without HFT the average transaction price is not observed as often as prices at the extremes.}
\label{fig:3}       
\end{figure}


The model reproduces several empirical findings that have otherwise been difficult to explain\cite{Hendershott, Hasbrouck, Brogaard}: 

\begin{enumerate}
\item Transaction prices are more accurate when HFT is present, i.e., they are closer to the equilibrium value.
\item Volatility is reduced when HFT is present.
\item Liquidity is increased when HFT is present.
\end{enumerate}

The equilibrium price, defined as the intersection of the expected aggregate supply and demand curve in the simulation, is just the mean of the uniform distribution of prices, i.e., 100.5. As seen in Table 1, the average transaction price both with and without HFT converges to the equilibrium value within the 2 standard error range that defines a 95\% confidence interval.  However, the variance around the equilibrium value is reduced when HFT is added.  The reduction in variance is shown in Table 1 and can be seen in the comparison of the histogram of transaction prices in Figures 3(A) and 3(B).  As seen in the figure, HFT causes more transactions to occur near the equilibrium price.  This result matches previous empirical studies that have shown algorithmic trading in general and HFT specifically increases the accuracy of prices in markets\cite{Brogaard, Hendershott}.

%
\begin{figure}[htb]
\includegraphics[scale=.30]{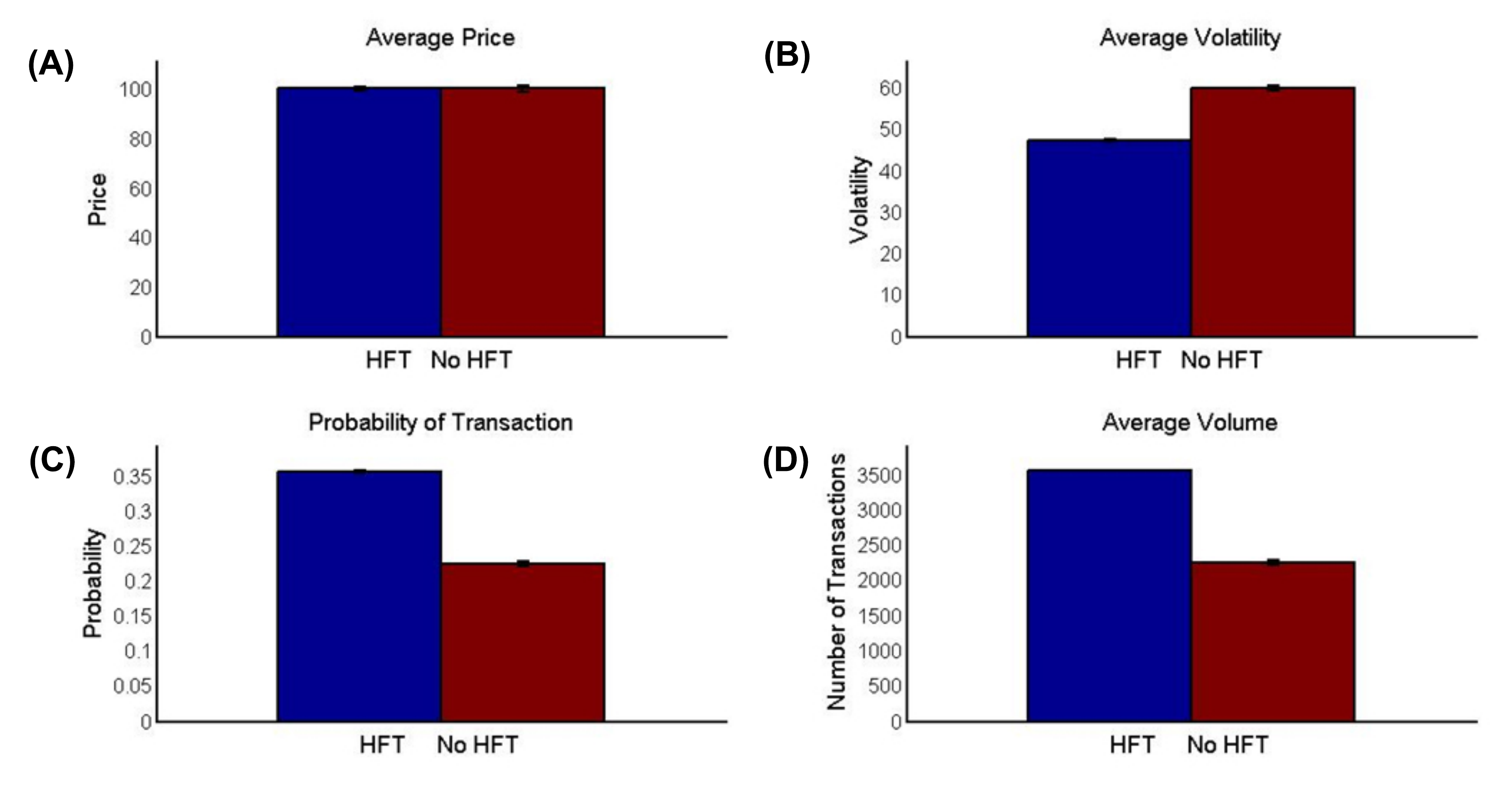}
%
%
\caption{Comparison plots of (A) average transaction price, (B) volatility, (C) transaction probability, and (D) volume (number of trades) in the simulation both with and without HFT.  Note that the introduction of HFT has no discernible effect on price, but statistically significant reduction of volatility, along with an increase in the number of trades and the likelihood of a given order being filled. Error bars denote 95\% confidence intervals.}
\label{fig:4}       
\end{figure}

Empirical studies have also found that HFT reduces intraday volatility\cite{Hasbrouck}.  Our simulation reproduces this result as well (see Figure 4(B)).  Because the equilibrium price is constant in the model, any variance in transaction price can be interpreted as excess volatility.  Because HFT reduces the variance of execution prices, it also reduces the volatility of the market.

The final metric we consider is liquidity. An asset is liquid if ``it is more certainly realizable at short notice without loss''\cite{Keynes}.  Liquidity can be defined quantitatively in a number of ways. However, our model accounts for the requirement of short notice, as orders are canceled if they do not result in a transaction, and when they do transact, the price must satisfy the reservation price initially generated. As a result, our measure of liquidity is the number of transactions that take place per simulation, or the probability that an order transacts. As shown in Table 1 and Figures 4(C and D), orders are more likely to be filled when HFT activity is present in our model.  Again, this result matches empirical findings\cite{Hasbrouck}.

%
\begin{table}
\caption{Average of parameters over 100 runs of 10000 iterations of the simulation.  Standard deviations are shown in parentheses below the average.  For the HFT case, the average is taken over both markets.}
\label{tab:1}       
%
%
\begin{tabular}{p{3.7cm}p{3.7cm}p{3.7cm}}
\hline\noalign{\smallskip}
Parameter & Base Case & HFT \\
\noalign{\smallskip}\svhline\noalign{\smallskip}
Price & 100.5 & 100.5 \\
 & (1.38) & (0.71)\\
Volatility & 60.1 & 47.4 \\
 & (0.6) & (0.4)\\
Volume & 2255 & 3580 \\
 & (33) & (6)\\
Probability of Transaction & 0.226 & 0.358 \\
\noalign{\smallskip}\hline\noalign{\smallskip}
\end{tabular}
\end{table}

\section{Conclusions}
\label{sec:5}

In this chapter, we analyzed the effects of high-frequency trading in a simulated environment.  With the premise that HFT activity connects orders across markets, we found that prices are closer to their equilibrium value, volatility is reduced, and liquidity is increased when HFT is present.  These results suggest that connecting order flow across similar securities is important for investor welfare, and to the extent that HFT performs this function, it serves an important purpose in modern financial markets.

%

\begin{acknowledgement}
This chapter is a modified version of Benjamin Myers MPhys thesis originally entitled ``Agent Based Simulations of High-Frequency Trading in Financial Markets.''  This work was supported by the European Commission FP7 FET-Open Project FOC-II (no. 255987). 
\end{acknowledgement}


\begin{thebibliography}{99.}

\bibitem{Brogaard} Brogaard J, Hendershott T, Riordan R (2013) High Frequency Trading and Price Discovery. Working paper, http://ssrn.com/abstract=1928510

\bibitem{Farmer} Farmer JD, Patelli P, Zovko II (2005) The Predictive Power of Zero Intelligence in Financial Markets. Proc Natl Acad Sci USA 102(6):2254-2259

\bibitem{Gerig2012} Gerig A (2012) High-Frequency Trading Synchronizes Prices in Financial Markets. Working paper, http://ssrn.com/abstract=2173247

\bibitem{GerigMichayluk2010} Gerig A, Michayluk D (2010) Automated Liquidity Provision and the Demise of Traditional Market Making. Working paper, http://ssrn.com/abstract=1639954

\bibitem{Gode} Gode DK, Sunder S (1993) Allocative Efficiency of Markets with Zero-Intelligence Traders: Market as a Partial Substitute for Individual Rationality. J Pol Econ, 101(1):119--137

\bibitem{Hasbrouck} Hasbrouck J, Saar G (2013) Low-Latency Trading. J Fin Markets 16(4):646-679

\bibitem{Hendershott} Hendershott T, Jones CM, Menkveld AJ (2011) Does Algorithmic Trading Improve Liquidity. J Fin 66(1):1--33

\bibitem{Keynes} Keynes JM (1930) A Treatise on Money. Vol 2: The Applied Theory of Money. Harcourt, New York

\bibitem{MacKenzie} MacKenzie D (2012) Mechanizing the Merc: The Chicago Mercantile Exchange and the Rise of High-Frequency Trading. Working paper, University of Edinburgh

\end{thebibliography}
\end{document}